# Spatial Localization Problem and the Circle of Apollonius.

Joseph Cox[1], Michael B. Partensky[2]

[1] Stream Consulting, Rialto Tower, Melbourne, Victoria 3000 Australia
email: joseph.jcox@gmail.com
[2]Brandeis University, Rabb School of Continuing Studies and Dept. of Chemistry, Waltham, MA, USA
email:partensky@gmail.com

The Circle of Apollonius is named after the ancient geometrician Apollonius of Perga. This beautiful geometric construct can be helpful when solving some general problems of mathematical physics, optics and electricity. Here we discuss its applications to "source localization" problems, e.g., pinpointing a radioactive source using a set of Geiger counters. The Circles of Apollonius help one to analyze these problems in a transparent and intuitive manner. This discussion can be useful for high school physics and math curriculums.

## 1. Introduction

We will discuss a class of problems where the position of an object is determined based on the analysis of some "physical signals" related to its location. First, we pose an entertaining problem that helps trigger the students' interest in the subject. Analyzing this problem. we introduce the Circles of Apollonius, and show that this geomteric insight allows solving the problem in an elegant and transparent way. At the same time, we demonstrate that the solution of the inverse problem of localizing an object based on readings from the detectors, can be nonunique. This ambiguity is further discussed for a typical "source localization" problem, such as pinpointing a radioctive source with a set of detectors. It is shown for the planar problem that the "false source" is the inverse point of the real one relative to the circle passing through a set of three detectors. This observation provides an insight leading to an unambiguous pinpointing of the source.

## 2. Apollonius of Perga helps to save Sam

### Description of the problem

Bartholomew the Frog with Precision Hopping Ability could hop anywhere in the world with a thought and a leap [1]. Publicly, he was a retired track and field star. Privately, he used his talent to help save the world. You see, Bartholomew had become a secret agent, a spy - a spook. In fact, only two people in the whole world knew who Bartholomew really was. One was Sam the Elephant and the other was Short Eddy, a fourteen-year-old kid who did not have a whole lot of normal friends but was superb in math and science. One day an evil villain Hrindar the platypus kidnapped Sam. Bartholomew, as soon as he realized Sam was missing, hopped straight "to Sam the Elephant." When he got there, he was shocked to see Sam chained to a ship anchored in the ocean. As soon as Sam saw Bartholomew he knew he was going to be okay. He quickly and quietly whispered, "Bartholomew, I don't exactly know where we are, but it is somewhere near Landport, Maine." It was dark out and Bartholomew could hardly see anything but the blurred outline of the city on his left, and the lights from three lighthouses. Two of them, say A and B, were on land, while the third one, C, was positioned on the large island. Using the photometer from his spy tool kit, Bartholomew found that their brightnesses were in proportion 36:9:4. He hopped to Eddy and told him what was up. Eddy immediately googled the map of the area surrounding Landport that showed three lighthouses (see Fig. 1). ABC turned out to be a right triangle, with its legs |AB|=1.5 miles and |AC|=2 miles. The accompanying description asserted that the lanterns on the lighthouses were the same. In a few minutes the friends knew the location of the boat, and in another half an hour, still under cover of the night, a group of commandos released Sam and captured the villain. *The question is, how did the friends manage to find the position of the boat?*

**Discussion and solution**

Being the best math and science student in his class, Eddy immediately figured out that the ratio of the apparent brightnesses could be transformed into the ratio of the distances. According to *the inverse square law*, the apparent brightness (intensity, luminance) of a point light source (a reasonable approximation when the dimensions of the source are small compared to the distance $r$ from it) is proportional to $P/r^2$, where $P$ is the power of the source. Given that all lanterns have equal power $P$, the ratio of the distances between the boat ("S" for Sam) and the lighthouses is

|SA|:|SB|:|SC|=1:2:3 (the square roots of 1/36, 1/9 and 1/4). Eddy always tried to break a complex problem into smaller parts. Therefore, he decided to focus on the two lighthouses, A and *B*, first. Apparently, *S* is one of *all possible points P* two times more distant from *B* than from *A*: $|PA|/|PB|=1/2$. This observation immediately reminded Eddy of something that had been discussed in his AP geometry class. At that time he was very surprised to learn that in addition to being the locus of points *P equally distant from a center*, a circle can also be defined *as a locus of points whose distances to two fixed points A and B are in a constant ratio*. Eddy opened his lecture notes and... There it was! The notes read: "Circle of Apollonius ... is the locus of points *P* whose distances to two fixed points *A* and *B* are in a constant ratio $\gamma$:

$$|PA|/|PB|=\gamma \qquad (1)$$

For convenience, draw the x-axis through the points A and B. It is a good exercise in algebra and geometry (see the Appendix) to prove that the radius of this circle is

$$R_0 = \gamma \frac{|AB|}{|\gamma^2 - 1|} \qquad (2)$$

and its center is at

$$x_O = \frac{\gamma^2 x_B - x_A}{\gamma^2 - 1} \qquad (3)$$

The examples of the Apollonius circles with the fixed points *A* and *B* corresponding to different values of $\gamma$ are shown in Fig. 2. Each of the Apollonius circles defined by Eq. 1 is the inversion circle [3] for the points *A* and *B* (in other words, it divides *AB harmonically*):

$$(x_A - x_O) \cdot (x_B - x_O) = R_O^2 \qquad (4)$$

This result immediately follows from Eqs. 2 and 3. (Apollonius of Perga [261-190 b.c.e.] was known to contemporaries as "The Great Geometer". Among his other achievements is the famous book *Conics* where he introduced such commonly used terms as parabola, ellipse and hyperbola [2])".

Equipped with this information, Eddy was able to draw the Apollonius circle $L_1$ for the points $A$ and $B$, satisfying the condition $\gamma = 1/2$ (Fig. 3). Given $|AB|=1.5$ and Eq. 2, he found that the radius of this circle $R_1 = 1$ mile. Using Eq. 3, he also found that $x_O - x_A = -0.5$ mile which implies that the center O of the circle $L_1$ is half a mile to the south from A. In the same manner Eddy built the Apollonius circle $L_2$ for the points $A$ and $C$, corresponding to the ratio $\gamma = |PA|/|PC|=1/3$. Its radius is $R_2 = 0.75$ mile and the center $Q$ is 0.25 mile to the West from $A$. Eddy put both circles on the map. Bartholomew was watching him, and holding his breath. "I got it!"- he suddenly shouted. "Sam must be located at the point that belongs simultaneously to both circles, i.e. right in their intersection. Only in this point his distance to *A* will be 2 times smaller than the distance to *B* and at the same time 3 times smaller than the distance to *C* ". "Exactly!"- responded Eddy, and he drew two dots, gray and orange. Now his friend was confused: "If there are two possible points, how are we supposed to know which one is the boat?" "That's easy"- Eddy smiled joyfully- "The gray dot is far inland which leaves us with only one possible location!". And Eddy drew a large bold "S" right next to the orange dot. Now it was peanuts to discover that the boat with poor Big Sam was anchored approximately 0.35 miles East and 0.45 miles North from A. Bartholomew immediately delivered this information to the commandos, and soon Big Sam was released. Once again, the knowledge of physics and math turned out to be very handy.

## 3. The question of ambiguity in some source localization problems

Our friends have noticed that the solution of their problem was not unique. The issue was luckily resolved, however, because the "fictitious" location happened to be inland. In general, such an ambiguity can cause a problem. Had both the intersection points appeared in the ocean, the evil villain would have had a 50:50 chance to escape. Thus, it is critical to learn how to deal with this ambiguity in order to pinpoint the real target and to discard false solutions.

We address this issue using a slightly different setting, quite typical for the localization problems. In the previous discussion a measuring tool, the photo detector, was positioned right on the object (the boat) while the physical signals used to pinpoint the boat were produced by the lanterns. More commonly, the signal source is the searched object itself, and the detectors are located in known positions outside the object. Practical examples are a radioactive source whose position must be determined using Geiger counters, or a light source detected by the light sensors. Assuming that the source and detectors are positioned in the same plane, there are three unknown parameters in the problem: two coordinates, and the intensity of the source *P*. One can suggest that using three detectors should be sufficient for finding all the unknowns. The corresponding solution, however, will not be unique: in addition to the real source, it will return a false source, similar to the gray dot found by Eddy and Bartholomew. Let us now discuss the nature of this ambiguity and possible remedies. Consider a source $S$ of power $P$ located at the point ($x_S, y_S$), and three isotropic detectors $D_k$ ($k$ =1, 2, 3) placed at the points ($x_k, y_k$) (see Fig. 4). The intensities $I_k$ sensed by the detectors are related to the source parameters through the inverse square law, leading to the system of three algebraic equations:

$$P/d_{s,k}^2 = I_k, \quad k=1,2,3. \quad (5)$$

Here $d_{s,k} = \sqrt{(x_S - x_k)^2 + (y_S - y_k)^2}$ is the distance between the $k$-th detector and the source.

Finding the source parameters based on the observed data (e.g. by solving Eqs. 5), is often called the "inverse problem". To address the question of ambiguity, we chose a more direct and intuitive approach, allowing a simple geometric relation between two (due to the non-uniqueness) solutions of Eqs. 5.

Treating the source as given, we use Eqs. 5 to generate the observables $I_k$ (usually, a much easier task than resolving the source based on the observations). Using the Circles of Apollonius, we show that another (*image* or *false*) source exists that exactly reproduces $I_k$ generated by the real source. Clearly, the existence of such an image signifies the non-uniqueness of the inverse problem. Finally, a simple geometric relation between the real and image sources will prompt a remedy for treating the ambiguity and pinpointing the source. To proceed, we first notice that *for any point A and any circle L, a second point B exists such that L is an Apollonius Circle with the fixed points A and B*.

This immediately follows from the observation that $B$ is the inverse point of $A$ (and vice versa) relative to an Apollonius circle with the fixed points $A$ and $B$ (see Eq. 4). In other words, obtaining $B$ by inverting $A$ in an arbitrary circle $L$, automatically turns $L$ into the Apollonius Circle for A and B.

Fig. 4 shows a circle $L$ passing through the three detectors. Inverting the source $S$ in $L$ produces the point $S'$. Its distance from the center of the circle $O$ follows from Eq. 4:

$$x_{S'} = R_O^2 / x_S \quad (6)$$

The corresponding parameter $\gamma$ is obtained by applying Eq. 1 to the point $P$ shown in Fig. 3:

$$\gamma = \frac{x_S - R}{R - x_{S'}} \quad (7)$$

As explained above, $L$ is the Circle of Apollonius with the fixed points $S$ and $S'$.

This observation is directly related to the question of ambiguity. From the definition of the Apollonius Circle, any chosen point on $L$ is exactly $\gamma$ times closer to $S'$ than it is to the real source $S$. In conjunction with the inverse square law it implies that a "false" source of the power $P' = P/\gamma^2$ placed in $S'$ would produce exactly the same intensity of radiation at all the points on the circle $L$ as does the real source $S$. Therefore, it is generally impossible to distinguish between the real and the false sources based on the readings from three (isotropic) detectors. Apparently, this is also true for any number of detectors *placed on the same circle*. This is exactly the reason for the ambiguity (nonuniqueness) of the inverse problem. Notorious for such ambiguities, the inverse problem is often characterized as being "ill-posed".

Eqs. 5 typically (*except for the case where S is placed right on L*) return two solutions, one for the real and the other for the false source. This ambiguity can be resolved by adding a fourth detector positioned off the circle *L*. Repeating the previous analysis for the second triad of detectors (e.g., 1, 3 and 4 positioned on the circle $L_1$, see Fig. 4), we can find a new pair of solutions: the original source *S* and its image $S''$. Comparing this with the previous result allows pinpointing the source *S*, which is the common solution obtained for the two triades of detectors, and filtering out the false solutions.

## 4. Conclusions

Using a simple geometric approach based on the Circles of Apollonius[*], we have shown that

(a) A planar isotropic (with three unknowns) source localization problem posed for a set of three detectors is typically non-unique.

(b) The "real" (*S*) and the "false" ($S'$) solutions are the mutually inverse points relative to the circle *L* through the detectors (the Apollonius circle for *S* and $S'$).

(c) Placing additional detectors on the same circle (e.g., in the vertexes of a polygon) does not help pinpoint the real source uniquely.

(d) With a fourth detector placed off the circle *L*, the real source can be found uniquely as a common solution obtained for two different sets of three detectors chosen out of four. Two other solutions (see $S'$ and $S''$ in Fig. 4) must be rejected.

Finally note that our analysis completely ignored the statistical fluctuations (noise) in the source and detectors, which is another important cause of ambiguity. Dealing with the noise usually requires additional detectors and special analytical methods (e.g., nonlinear regression). Nevertheless, the geometric ideas described above can still be useful in these applications.

## 5. Appendix

With the x-axis passing through A and B (see Fig. 3), the coordinates of these points are correspondingly $(x_A,0)$ and $(x_B,0)$. Let $(x,y)$ be the coordinates of a point *P* satisfying Eq. 2. Squaring Eq.1 and expressing |PA| and |PB| through the coordinates we find:

$$(x - x_A)^2 + y^2 = \gamma^2[(x - x_B)^2 + y^2] \quad (A1)$$

Expanding the squares, dividing by $1 - \gamma^2$ (the case $\gamma = 1$ is discussed separately) and performing some simple manipulations, we can derive the following equation:

$$(x - x_O)^2 + y^2 = R_O^2 \quad (A2)$$

with

$$R_O = \left| \frac{\gamma(x_A - x_B)}{\gamma^2 - 1} \right|, \quad x_O = \frac{x_A - \gamma^2 x_B}{1 - \gamma^2} \quad (A3)$$

Eq. A2 describes the circle of radius $R_O$, with its center at $x_O$. Eqs. A3 are equivalent to Eqs. 2 and 3, which proves the validity of those equations.

The solution for $\gamma = 1$ obtained directly from Eq. A1 is the straight line perpendicular to *AB* and equidistant from the points *A* and *B*.

___________________________________

[*] Some similar geometric ideas also inspired by Apollonius of Perga, are discussed in ref. [4] in application to GPS.

## Acknowledgements

We are grateful to Jordan Lee Wagner for helpful and insightful discussion. MBP is thankfull to Arkady Pittel and Sergey Liberman for introducing him to some aspects of the source localization problem, and to Lee Kamentsky, Kevin Green, James Carr and Philip Backman for valuable comments and suggestions.

# Figures

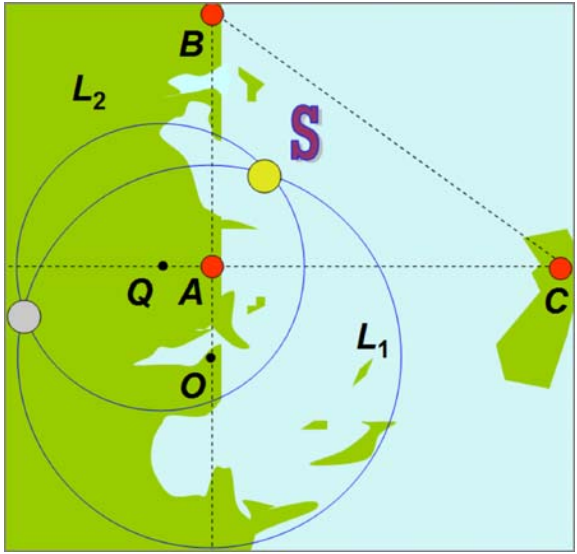

**Fig.1** The map of the Landport area showing three lighthouses marked A, B and C. Other notations are explained in the text.

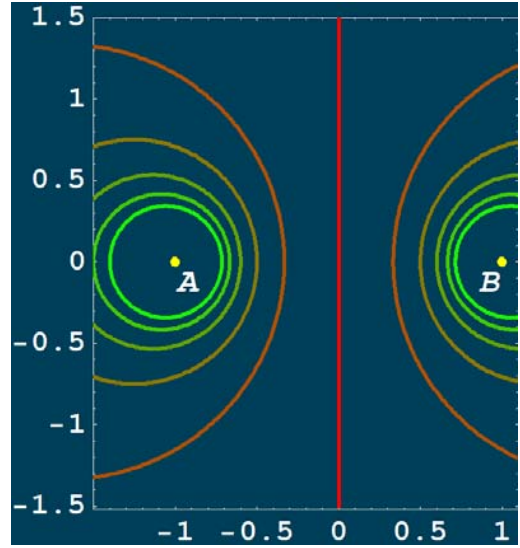

**Fig. 2**: The Circles of Apollonius (some are truncated) for the points A(-1,0) and B(1,0) corresponding to ratio $\gamma=k$ (right) and $\gamma=1/k$ (left), with $k$ taking integer values from 1 (red straight line) through 6 (bright green).

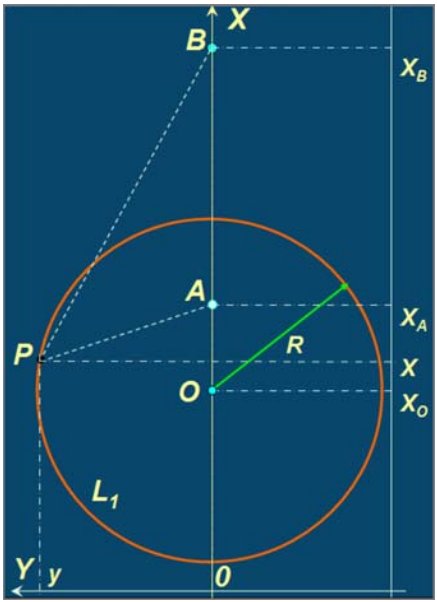

**Fig. 3**: Construction of the Apollonius circle $L_1$ for the points $A$ and $B$. Distance $|AB| = 1.5$, $R=1$, $|OA|= 0.5$ (miles). For any point $P$ on the circle, $|PA|/|PB| =1/2$. It is clear from the text that the lantern $A$ looks from $P$ four times brighter than $B$. The x-coordinates of the points $A$, $B$ and $O$ are shown relative to the arbitrary origin $x =0$. Note that only the ratio of brightnesses is fixed on the circle while their absolute values vary.

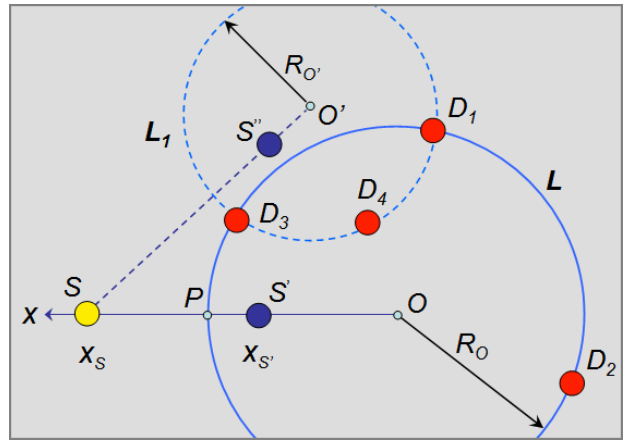

**Fig. 4** Pinpointing the source $S$. $L$ is the circle through the first three detectors; it is the Apollonius Circle for the original source $S$ and the false source $S'$ (two sulutions of the inverse problem). The detector $D_4$ is positioned off $L$. The circle $L_1$ passes through the detectors 1, 3 and 4. The corresponding solutions are $S$ and $S''$.